 \documentclass[authoryear,final,twocolumn,12pt]{elsarticle}

\usepackage{epsfig}
\usepackage{rotating}
\usepackage{journalnames}
\usepackage{color}
\usepackage[none]{hyphenat}
\usepackage{amssymb}

\journal{Icarus}
\begin{document}

\sloppy

\begin{frontmatter}



\title{The Spherical Brazil Nut Effect and its Significance to Asteroids}

\author[ASU]{Viranga~Perera\corref{cor1}}
\ead{viranga@asu.edu}
\cortext[cor1]{Corresponding author} 

\author[ASU]{Alan~P.~Jackson}
\author[ASU]{Erik~Asphaug}
\author[UM]{Ronald-Louis~Ballouz}

\address[ASU]{School of Earth and Space Exploration, Arizona State University, PO Box 876004, Tempe, AZ 85287-6004, USA.} 
\address[UM]{Department of Astronomy, University of Maryland, 1113 Physical Sciences Complex, Bldg. 415, College Park, MD 20742-2421, USA.}

\begin{abstract}

Many asteroids are likely rubble-piles that are a collection of smaller objects held together by gravity and possibly cohesion. These asteroids are seismically shaken by impacts, which leads to excitation of their constituent particles. As a result it has been suggested that their surfaces and sub-surface interiors may be governed by a size sorting mechanism known as the Brazil Nut Effect. We study the behavior of a model asteroid that is a spherical, self-gravitating aggregate with a binary size-distribution of particles under the action of applied seismic shaking. We find that above a seismic threshold, larger particles rise to the surface when friction is present, in agreement with previous studies that focussed on cylindrical and rectangular box configurations. Unlike previous works we also find that size sorting takes place even with zero friction, though the presence of friction does aid the sorting process above the seismic threshold. Additionally we find that while strong size sorting can take place near the surface, the innermost regions remain unsorted under even the most vigorous shaking.

\end{abstract}

\begin{keyword}

Asteroids \sep Asteroids, surfaces \sep Interiors




\end{keyword}

\end{frontmatter}


\section{Introduction}

Asteroids are small bodies that are remnants of the early planet formation process \citep{Asphaug_2009}. Space missions have imaged certain asteroids and as a result have greatly helped the understanding of asteroid surface properties. However, due to the lack of seismic data, it has been difficult to definitively constrain the internal structure of asteroids. The understanding of their internal structures is important for planetary science, for future asteroid exploration and mining \citep{Hatch_2015}, and for deterring potential Earth impact hazards \citep{NRC_2010}.

Previous works have inferred that asteroids 150~m to 10~km in size are likely rubble-pile objects that are a collection of smaller objects held together by gravity and possibly cohesion \citep{Michel_2001, Richardson_2002, Pravec_2002, Sanchez_2014}. This characterization arises from several key observations: 
\begin{enumerate}
\item Craters on their surfaces and the dynamical evolution of asteroids indicate that asteroids have undergone many impacts over their lifetimes that will have left disrupted, reaccumulated objects \citep{Asphaug_1998, Richardson_2004}. 

\item Low bulk densities and high macroporosities of asteroids indicate the presence of large internal voids \citep{Carry_2012}. 

\item The limited spin rates of asteroids possibly point to loosely held aggregates \citep{Scheeres_2015}. 

\item Spacecraft images have shown that some asteroids have large boulders that seem to be protruding from their surfaces such as Eros \citep{Asphaug_2001} and Itokawa \citep{Miyamoto_2007, Tancredi_2015}.
\end{enumerate}
 
As a rubble-pile asteroid is being seismically shaken by impacts, its constituent particles should undergo granular flow once frictional forces are overcome. Particularly, the Brazil Nut Effect where larger constituent objects rise to the top against gravity may occur on these rubble-pile asteroids (assuming the constituent objects are approximately the same density). Past work has shown that when a collection of particles of varying sizes is excited, over time larger particles will accumulate at the top given gravity is downward \citep{Rosato_1987}. Some large boulders on asteroids could be the result of the Brazil Nut Effect, though this may not be the only mechanism for producing large surface boulders \citep[e.g.][]{Thomas_2001}.

The Brazil Nut Effect is a complex phenomenon, but it has been proposed to be mediated through two primary mechanisms:
\begin{enumerate}
\item Smaller particles may fill in and pass through spaces created by excitations while larger particles do not \citep{Williams_1976}. If the direction of gravity is downward, this results in smaller particles migrating to the bottom while larger particles are ratcheted upwards.

\item Depending on boundary conditions, excitation of particles may set up granular convection that brings larger particles to the top but prevents them from moving downward \citep{Knight_1993}.
\end{enumerate}

The Brazil Nut Effect has been studied in a terrestrial context through computer simulations using hard spheres (i.e. simulated spheres do not deform when forces are applied to them) \citep{Rosato_1987}, using soft spheres (i.e. simulated spheres deform when forces are applied to them) \citep{Kohl_2014}, and through experiments in cylindrical columns \citep{Knight_1993}. Additionally, in the context of the low-gravity environments of asteroids, simulations have been done using a soft spheres method in rectangular and cylindrical box configurations \citep{Tancredi_2012, Matsumura_2014} and parabolic flight experiments have been done in a cylindrical configuration to momentarily obtain equivalent low-gravity conditions of the Moon and Mars \citep{Guttler_2013}. 

Previous preliminary work in two-dimensions has suggested that size sorting can occur in self-gravitating aggregates \citep{Sanchez_2010}; however, the Brazil Nut Effect has not been studied in a fully three-dimensional configuration. Here we have conducted simulations using a spherical, self-gravitating configuration of particles since that configuration is more representative of asteroids. In Section \ref{sec:Method}, we discuss the $N$-body gravity code we used (Section \ref{sec:Method:pkdgrav}) and our initial conditions along with a short discussion of how we created the aggregate that was used for the simulations (Section \ref{sec:Method:InitialConditions}). In Section \ref{sec:Method:Simulations} we describe the simulations that were conducted and the section concludes with a discussion of how we compare our simulations to asteroids (Section \ref{sec:Method:CompareAsteroids}). In Section \ref{sec:Results}, we state our results while focussing on the central region of the aggregate (Section \ref{sec:Results:CentralRegion}), the effect of friction (Section \ref{sec:Results:Friction}), and the time evolution of particle distributions (Section \ref{sec:Results:TimeEvolution}). In Section \ref{sec:Discussion}, we discuss our results considering asteroid surface processes (Section \ref{sec:Discussion:AsteroidSurfaces}), the central region of our aggregate (Section \ref{sec:Discussion:CentralRegion}), and the driving mechanism of the Brazil Nut Effect (Section \ref{sec:Discussion:DrivingMechanism}). Finally, we summarize and discuss future work in Section \ref{sec:Summary}.

\section{Method}
\label{sec:Method}

\subsection{\textsc{pkdgrav}}
\label{sec:Method:pkdgrav}

For our work we used \textsc{pkdgrav}, a parallel $N$-body gravity tree code \citep{Stadel_2001} that has been adapted for particle collisions \citep{Richardson_2000, Richardson_2009, Richardson_2011}. Originally collisions in \textsc{pkdgrav} were treated as idealized single-point-of-contact impacts between rigid spheres. We use a soft-sphere discrete element method (SSDEM) to model the collisions of particles. In SSDEM, particles are allowed to slightly overlap with one another. Particle contacts can last many time steps, with reaction forces dependent on the degree of overlap (a proxy for surface deformation) and contact history. The code uses a second-order leapfrog integrator to solve the equations of motion, with accelerations due to gravity and contact forces recomputed each step.

The spring/dash-pot model used in \textsc{pkdgrav}'s soft-sphere implementation is described fully in \citet{Schwartz_2012} and is based on \citet{Cundall_1979}. Two overlapping particles feel a Hooke's law type reaction force in the normal and tangential directions determined by spring constants ($k_n$ and $k_t$). We chose a normal spring constant ($k_n$) that kept particle overlaps \textless 1\%. The choice of a linear spring was made during the original implementation of the soft-sphere code. While a Hertzian spring contact may provide benefits in certain circumstances we find the linear spring to be adequate for the problem at hand, with the added advantage of simplicity. In particular, we note that experimentally the coefficient of restitution of meter-scale granite spheres has been found to have no dependence on impact speed for low-speed impacts \citep{Durda_2011}, which is suggestive of a linear contact response. User-defined normal and tangential coefficients of restitution used in hard-sphere implementations, $\epsilon_{n}$ and $\epsilon_{t}$, determine the plastic damping parameters ($C_n$ and $C_t$), which are required to resolve a soft-sphere collision (see Eq. 15 in \citealt{Schwartz_2012}). Frictional forces can also be imposed on the interaction by adjusting static, twisting, and rolling coefficients.

This SSDEM implementation has been validated through comparison with laboratory experiments (e.g., \citet{Schwartz_2012} and \citet{Schwartz_2013}). In addition, \citet{Ballouz_2015} used this SSDEM to model the collisions of rubble-pile asteroids made up of 40~m spheres, and showed that the outcomes of binary collisions were consistent with scaling laws for low- and high-speed collisions. Furthermore, \citet{Matsumura_2014} studied the classical Brazil Nut Effect for centimeter-sized grains in a cylindrical container using this method.

\subsection{Initial Conditions}
\label{sec:Method:InitialConditions}

The initial spherical aggregate used in the following simulations was made by creating 500 particles of radius 40~m (colored yellow) and 500 particles of radius 80~m (colored red) that were randomly positioned inside a cubic space of 4~km per side. All particles had a density of 3~g/cm\textsuperscript{3}. Particles were then allowed to gravitationally collapse due to self-gravity with the coefficients of friction set to zero to form a mixed aggregate and left to settle for 75 simulation hours. The maximum free-fall time of the initial cubic distribution of particles (i.e. from the corners) is around 3 hours.

The aggregate that was created in the process had a mass of $3.62 \times 10^{12} $ kg, a bulk radius of about 800~m, and a bulk density of about 1.7~g/cm\textsuperscript{3}. The aggregate properties are representative of common asteroids. The escape speed of the aggregate was 75~cm/s. In order to properly resolve particle collisions, we use a normal spring constant of \begin{equation} k_n = m_p (\frac{v_{\mathrm{max}}}{x_{\mathrm{max}}})^2\end{equation} where $m_p$ is the typical particle mass, $v_{max}$ is the maximum expected particle speed, and $x_{max}$ is the maximum expected fractional overlap, which we set to 1\% of the typical particle radius. This chosen value of $k_n$ allows all the kinetic energy of the particle collision to be stored in a single spring that compresses to $x_{\mathrm{max}}$. Furthermore, in order to ensure that a collision is properly resolved, we require that particle overlaps last at least 12 time steps for the smallest particles. The length of a single time step can be estimated by considering the oscillation half-period of a spring with normal spring constant $k_n$ (see Eq. 36--38 in \citealt{Schwartz_2012}). Using the typical particle sizes, masses, and expected speeds we find that a spring constant of $k_n \sim 4.856 \times 10^9$ kg/s$^2$ and a time step of $8.523 \times 10^{-2}$ s are required to properly resolve the collisions in our simulations. The tangential spring constant, $k_t$, is taken to be equal to $\frac{2}{7} \times k_n$. Tests with one half and one quarter of our chosen time step showed no deviation in behavior demonstrating that our chosen time step is adequate.

Since \citet{Matsumura_2014} found that the Brazil Nut Effect is largely insensitive to the choice of the coefficients of restitution and since we wanted to focus on the magnitude of seismic shaking and the coefficients of friction, we set the normal coefficient of restitution to 0.2 and the tangential coefficient of restitution to 0.5 for all our simulations. We will further examine the effect of these damping coefficients on the Brazil Nut Effect in a future study. 

\begin{figure}
\includegraphics[width=\columnwidth]{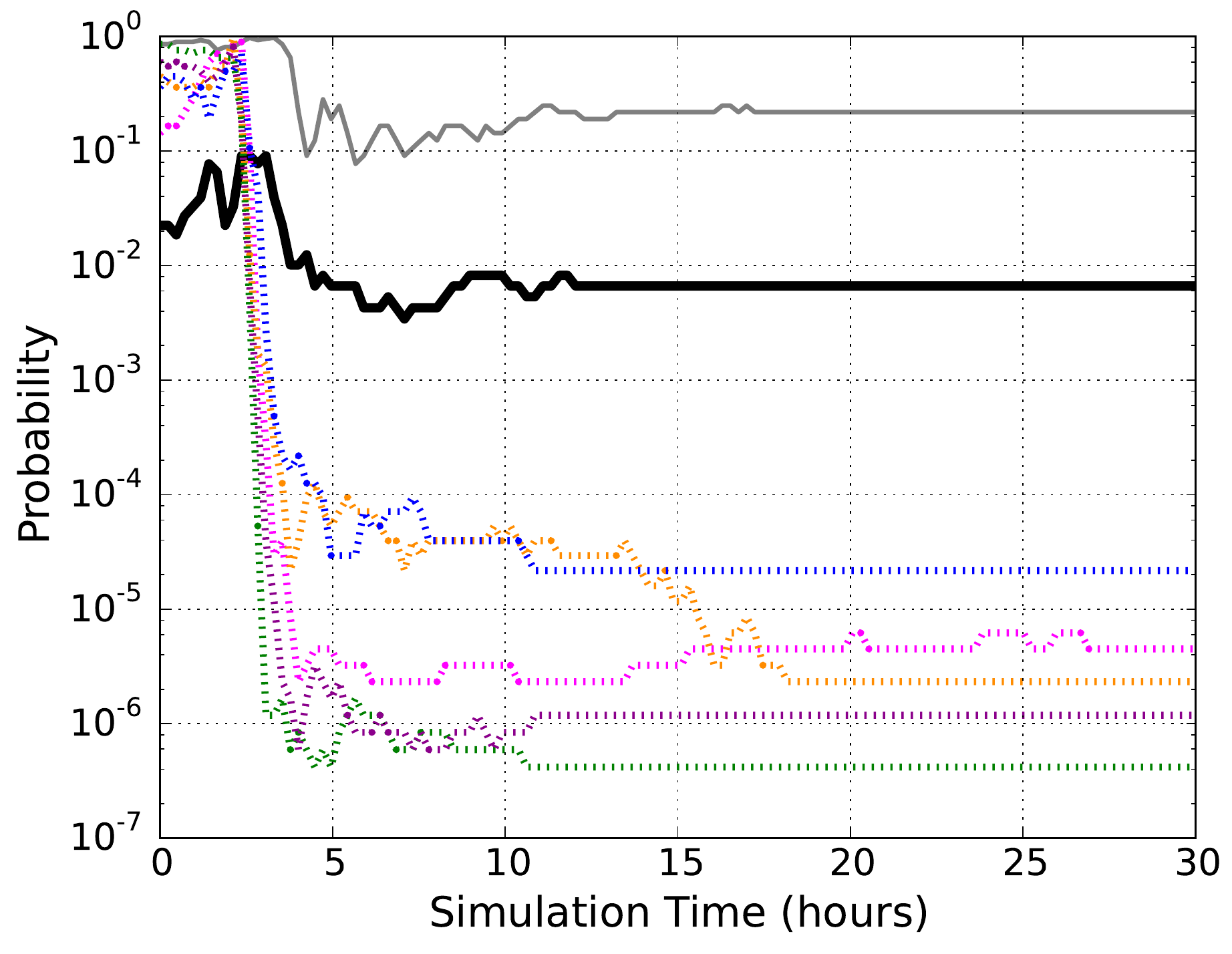}
\centering
\caption{Probabilities that larger (red) and smaller (yellow) particles were drawn from the same parent distribution for seven initial aggregates. Probabilities shown as a function of initial aggregate formation/settling time and were determined using the K-S test. For all seven initial aggregates the friction coefficients were set to zero (to ensure aggregates that were formed were spherical) and the coefficients of restitution were 0.2 and 0.5 for the normal and tangential directions respectively. The solid black line shows values for the aggregate composed of particles of two sizes used for this work. Colored dotted lines show other trial aggregates composed of particles of two sizes. The solid gray line shows an aggregate that was composed of particles of the same size but were randomly assigned either a color of red or yellow for comparison.}
\label{fig:KSplots}
\end{figure}

In Figure \ref{fig:KSplots} we show the likelihood that radial distributions of larger (red) and smaller (yellow) particles were drawn from the same parent population as a function of settling time.  A higher probability indicates that it is more likely that the two particle groups were drawn from the same parent population, and thus that their radial distributions are more similar to each other.  Probabilities were calculated using a two-sample Kolmogorov-Smirnov (K-S) test. The two-sample K-S statistic quantifies the distance between the cumulative distributions of the two samples (here the number of particles within a radius $r$), which determines the probability that the two samples are drawn from the same underlying distribution. The utility of the K-S test lies in the fact that it is a non-parametric test, and so allows us to make no assumptions about the shape of the underlying distribution. The thick black line shows the initial aggregate used in this study.  We also show five additional two-size aggregates (dashed lines) and an aggregate with equal-sized particles (solid gray line). We can see that the trial aggregates have largely settled by 5 hours and all have completely stabilized by 30 hours.

Since the particles in the equal-size case are identical aside from a randomly assigned color, we should expect the colors to be well mixed and so we can use this as the ideal case. By comparison it is clear that the aggregate used for this study (solid black line) displays a statistically significant difference between the two particle groups. Nevertheless, the aggregate used for this study shows less of a difference than the other two-sized particle aggregates (dashed colored lines), which enabled us to more easily distinguish further Brazil Nut Effect size sorting during our later simulations.

We attribute the statistically significant difference to the occurrence of the Brazil Nut Effect during the formation of the two-sized particle aggregates. When particles collapsed due to self-gravity during the formation of the aggregate, their mutual kinetic energies imparted a seismic shock that size sorted the particles. During the collapse process, the coefficients of friction were set to zero to ensure that the aggregate would be approximately spherical in shape. Using higher coefficients of friction could possibly reduce the effect of the initial size sorting; however, it would also introduce the issue of a misshaped aggregate that would complicate the analysis of the motion of the particle distributions. An alternative method would be to create the aggregate particle by particle; however, that would have been very computationally intensive.

\subsection{Simulations}
\label{sec:Method:Simulations}

All simulations started with the same original aggregate (black line in Figure \ref{fig:KSplots}). This ensured that all of our numerical results began from the same initial distributions, aiding comparison of the results. The parameters varied for this study were the magnitude of the shaking and the coefficients of friction. In the first set of simulations particles did not have friction (i.e. friction coefficients were set to zero) and in the second set of simulations we assigned each particle a static friction coefficient of 0.7 and a rolling friction coefficient of 0.1 similar to nominal values used in \citet{Matsumura_2014}. They showed that the set of friction parameters listed above would lead to the Brazil Nut Effect occurring for a range of seismic and gravitational environments in a cylindrical box configuration.

At the beginning of each simulation each particle was independently assigned a random velocity drawn from a distribution of velocities that ranged from 0 to $v_{\rm max}$. The values of $v_{\rm max}$ prescribed for the run are given in Table \ref{table:tableParameters}. The directions of the velocities were isotropically distributed with respect to the particle. After this `shaking,' particles were allowed to gravitationally settle for a period of 4.7 simulation hours (200,000 time steps). We can see from Figure \ref{fig:KSplots} that $\sim$5 hours is sufficient for the aggregate to have largely settled after the initial collapse and that the initial collapse is more violent than any individual shake in any of our simulations as well as having a longer lead time before settling can begin ($\sim$3 hours). After settling, all particles were again assigned new random velocities that were again no larger than the predefined maximum magnitude. For each run, the `shaking'€™ and settling process was repeated for 102 simulation days for a total of 516 `shakes' to mimic a prolonged period of seismic shaking. Each simulation run took approximately 15 days to complete.

For each of the friction and no friction sets, there were six simulations each for six different maximum magnitudes of velocity. Six speeds were chosen initially to be a percentage of the aggregate's estimated escape speed (1\%, 10\%, 25\%, 30\%, 40\%, and 50\%). These speeds were later converted to be a percentage of the aggregate's true escape speed (0.92\%, 9.24\%, 23.10\%, 27.73\%, 36.97\%, and 46.21\%). These values have no special significance other than to have a range of speeds to explore the parameter space. Friction coefficients and maximum magnitudes of velocity used for each run are listed in Table \ref{table:tableParameters}.

\begin{table*}
\caption{Simulation runs}
\centering
\begin{tabular}{l c c c c}
 Run & Static Friction & Rolling Friction & Max Speed (cm/s) & Max Speed (escape speed)\\
\hline
\hline
  1  & 0 & 0 & 0.692 & 0.92\% \\
  2  & 0 & 0 & 6.95 & 9.24\% \\
  3  & 0 & 0 & 17.4 & 23.10\% \\
  4  & 0 & 0 & 20.9 & 27.73\% \\
  5  & 0 & 0 & 27.8 & 36.97\% \\
  6  & 0 & 0 & 34.7 & 46.21\% \\
\hline
  7  & 0.7 & 0.1 & 0.692 & 0.92\% \\
  8  & 0.7 & 0.1 & 6.95 & 9.24\% \\
  9  & 0.7 & 0.1 & 17.4 & 23.10\% \\
  10  & 0.7 & 0.1 & 20.9 & 27.73\% \\
  11  & 0.7 & 0.1 & 27.8 & 36.97\% \\
  12  & 0.7 & 0.1 & 34.7 & 46.21\% \\
\hline
\label{table:tableParameters}
\end{tabular}
\end{table*}

\subsection{Considerations for comparisons with asteroids}
\label{sec:Method:CompareAsteroids}

There are several aspects we must consider when comparing our simulations to asteroids. Rubble-pile asteroids are believed to be composed of self-gravitating particles and have friction and restitution, much like our simulated aggregates. In a rubble-pile asteroid however the constituent particles will be non-spherical and the friction and restitution parameters will likely be complex. To a certain degree, we can consider the asphericity of the particles as a source of large scale friction due to interlocking, and so this can be partly accounted for by the coefficient of friction.

While asteroids, like our aggregates, are self-gravitating, they are typically not spherical, but rather have a variety of odd shapes \citep{Durech_2015}. While our simulations may not be exactly representative of the typical non-spherical asteroid, the important aspect lies in the fact that our aggregates are three-dimensional and self-gravitating. A corollary to this, which we believe is important, is that unlike simulations that involve box configurations our simulations do not have walls. Therefore, any size sorting that takes place can only be due to particle-particle interactions, and not due to particle-wall interactions.

\subsubsection{The size distribution}
\label{sec:Method:CompareAsteroids:SizeDistribution}

An important aspect for comparing our simulations (or any other model of the Brazil Nut Effect) with asteroids is the size distribution. The constituent particles of a rubble-pile asteroid will form part of a likely largely continuous size distribution, however the shape of this distribution is poorly understood. As such it is preferable to adopt a simple assumption that allows us to make inferences about the behavior without being reliant on highly uncertain details of the size distribution. The simplest such assumption is that of a binary size distribution with two populations of particles. Like many works before them, \citet{Matsumura_2014} focused on the intruder model. In intruder models, there is only one or a few large particle(s) in comparison to the quantity of small particles. \citet{Matsumura_2014} argue that since the internal structures of asteroids are poorly understood, that assumption is valid. Though the intruder model might be applicable in the case of a single or a few large boulders buried beneath the surface of an asteroid, it is not clear that the model is suitable to study granular flow of all constituent particles of an asteroid. Rather than an intruder model, we chose to adopt a different implementation of a binary size distribution, with equal numbers of large and small particles, which we believe may be a better representation of a bulk asteroid.

\subsubsection{The shaking model}
\label{sec:Method:CompareAsteroids:ShakeModel}

The main seismic input for most asteroids is most likely from impacts, though other sources are possible, such as unloading of tidal stresses during close-encounters. In an impact the seismic impulse will have a discrete source on the surface of the asteroid from which the seismic waves will propagate outwards, attenuating with distance travelled. This is clearly rather different from the method we use here, in which the seismic impulse is equally applied throughout the body. One important point here is that while an individual impact is a localized source we are not interested here in the effects of a single impact, but rather in the collective effect of many impacts over time. Individual impacts will occur at random locations on the surface, as such the bulk effect of many impacts over time will be uniformly distributed across the surface. In the interior we note that the seismic shock waves will reflect off the far side of the asteroid and off interior flaws within the body. Since asteroids are irregularly shaped these reflections will be chaotic and will likely lead to unpredictable foci and dead zones. As such it is unclear how attenuation into the interior of the asteroid should be handled, and so we choose to use our model of applying the seismic impulse uniformly throughout the interior for simplicity. We also note than any non-impact source of seismic disturbances, such as unloading of tidal stresses, would likely result in a more distributed source located below the surface, and that for any given shaking velocity our shaking model can be expected to give the maximum effect in the interior.

\section{Results}
\label{sec:Results}

\begin{figure*}
\includegraphics[width=1.0\textwidth]{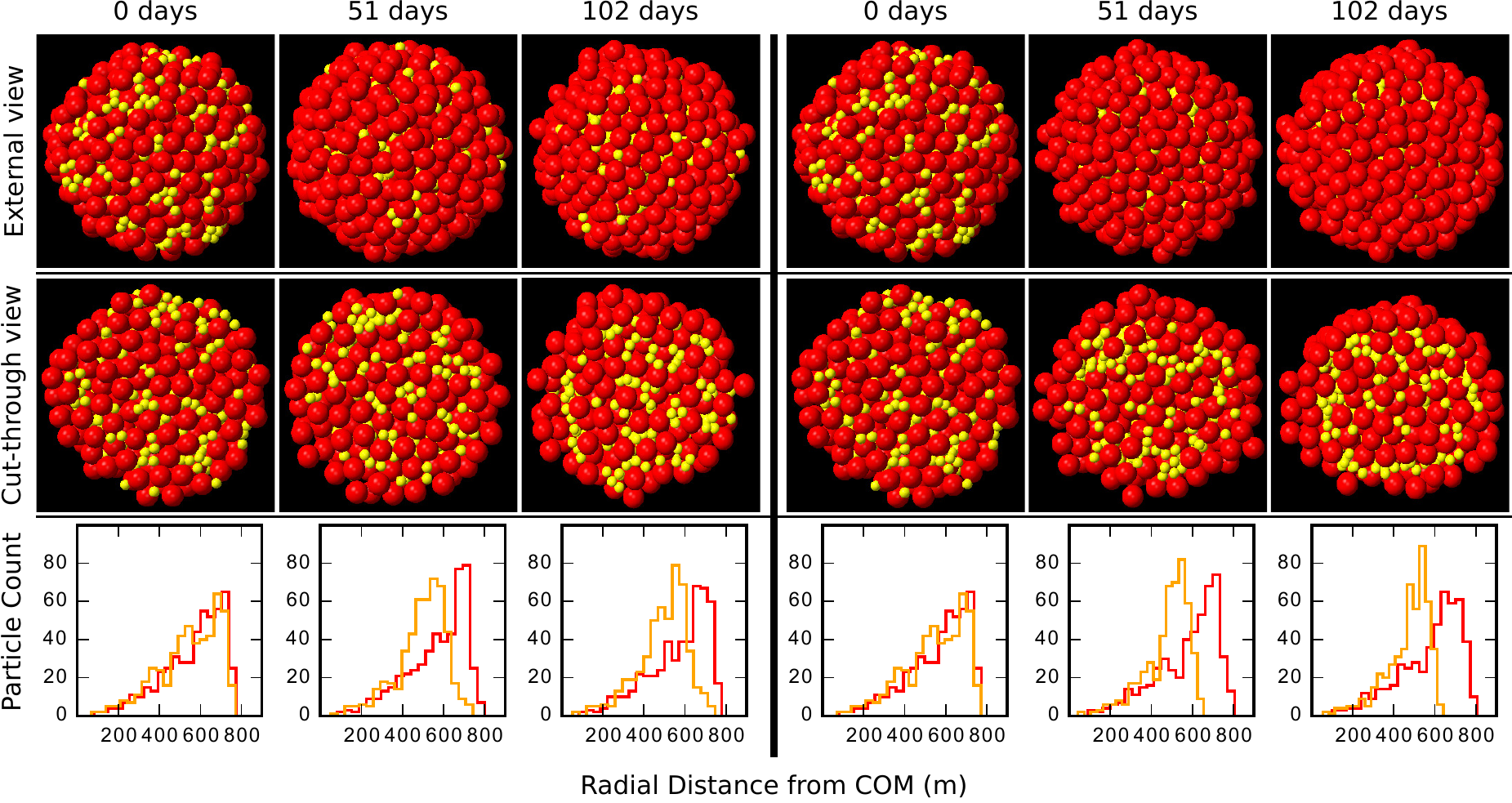}
\caption{Two simulation runs shown without friction on the left and with friction on the right. Larger particles (radius 80~m) are colored red and the smaller particles (radius 40~m) are colored yellow. For both the runs (i.e. Runs 6 and 12), the maximum magnitude of shaking was 46.21\% of the escape speed (34.7~cm/s). Each panel shows three stages (0, 51, and 102 days) of the simulations. Top row: external views. Middle row: cut-through views. Bottom row: histograms using a radius bin size of 20~m where particle radial distance is measure from the aggregate'™s center of mass. The yellow and red curves represent the smaller and larger particles respectively.}
\label{fig:aggregates_and_cutouts}
\end{figure*}

In Figure \ref{fig:aggregates_and_cutouts}, we show three time steps from Run 6 (no friction) on the left and three time steps from Run 12 (with friction) on the right. Progressive random shaking, in this case with a maximum magnitude of 46.21\% of the escape speed (34.7~cm/s), resulted in the mixed aggregate becoming sorted. Over time, larger (red) particles can be observed rising to the surface while smaller (yellow) particles that were on the surface submerged. In the final cut-through views, there are larger particles on the surfaces with smaller particles beneath them. The innermost regions remain well mixed however. These size sorted and well mixed regions are also discernible on the histograms of particle radial distance (bottom panels of Figure \ref{fig:aggregates_and_cutouts}).

\subsection{The well mixed central region}
\label{sec:Results:CentralRegion}

To explore this well mixed region further, we divided our aggregate into ten shells of 100 particles each. The first shell consisted of the first 100 particles from the center of mass, the second shell the next 100 particles and so forth. We chose to define the shells in this manner, rather than for example by defining them according to fixed radii, as this ensured that the shells always contained the same number of particles. This made comparisons between the simulation runs more straightforward.

\begin{figure*}
\centering
\includegraphics[width=1.00\textwidth]{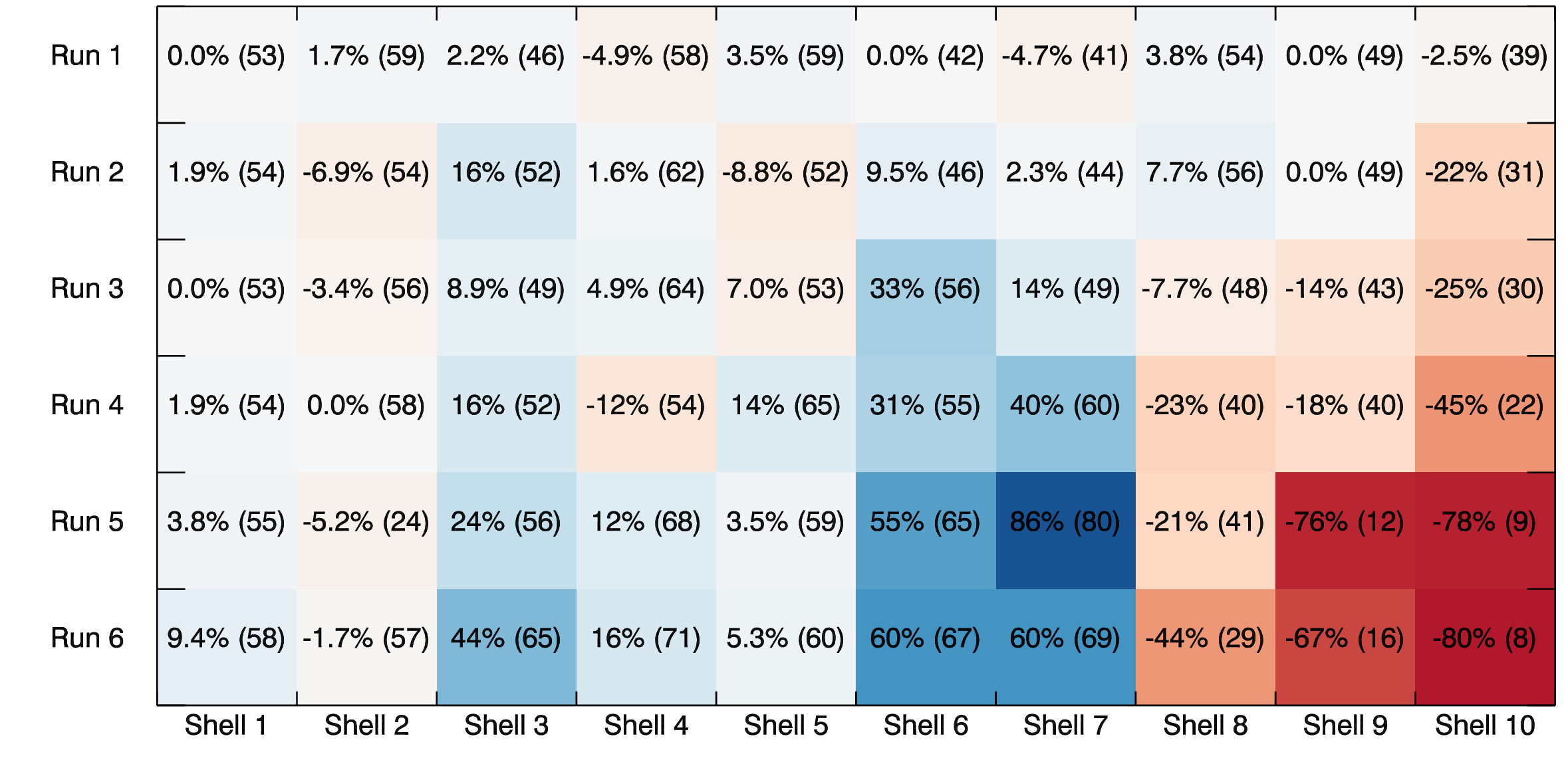}
\includegraphics[width=1.00\textwidth]{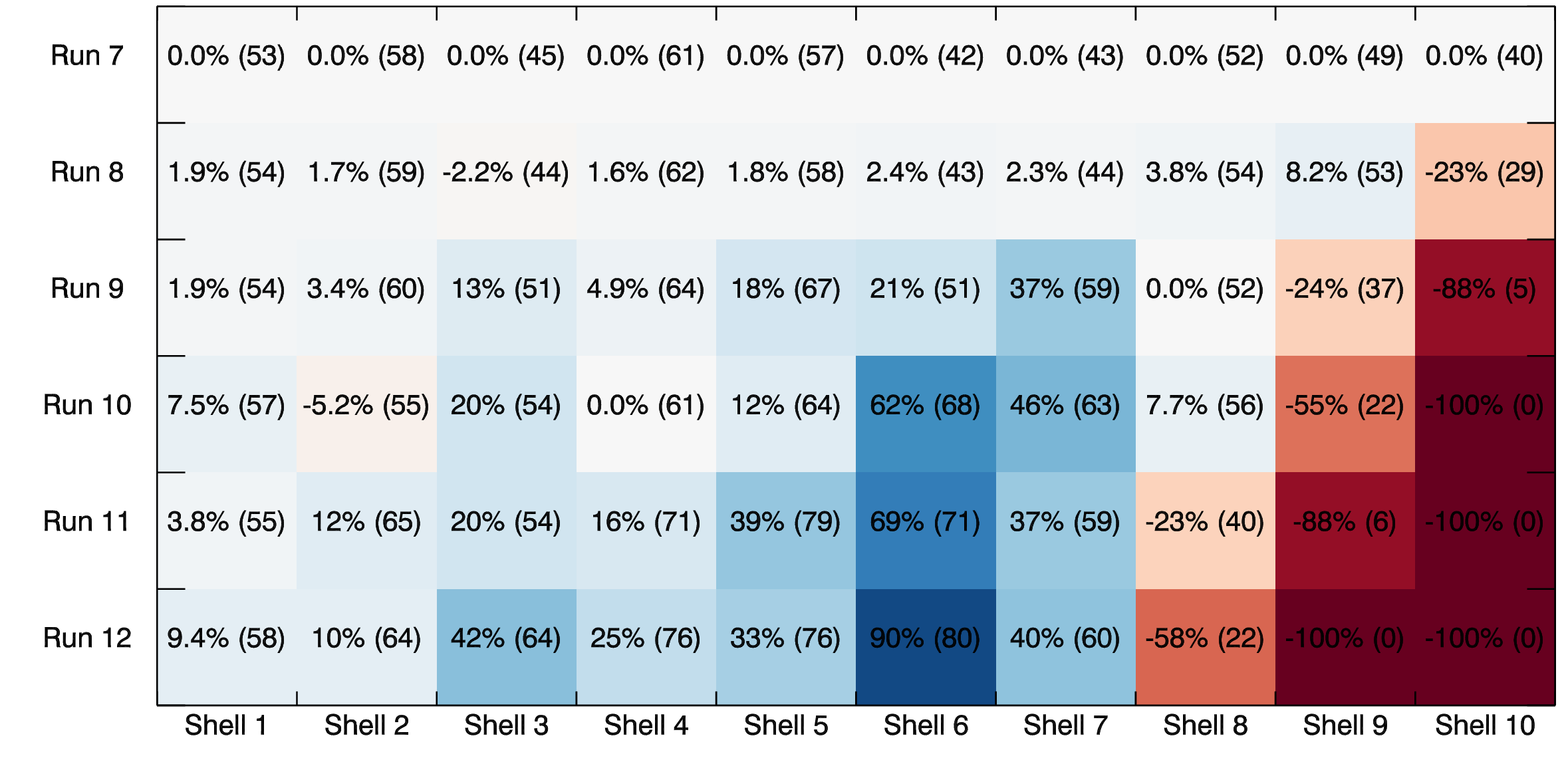}
\includegraphics[width=0.85\textwidth]{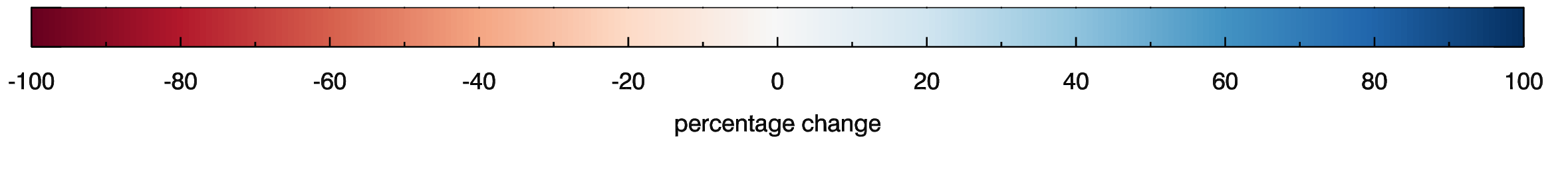}
\caption{Percentage change of the number of smaller (yellow) particles present inside defined spherical shells from beginning to end of the simulations. Shell 1 contains the first 100 particles from the center of the aggregate while each of the following shells has the next 100 particles. Shell 10 contains the last 100 particles from the center of the aggregate. The coloring indicates whether smaller particles are being depleted (red coloring) or whether they are being augmented (blue coloring). Number of smaller (yellow) particles present at the end of the simulations are listed in parenthesis for each shell. Runs 1 to 6 have no friction (top) and Runs 7 to 12 have friction (bottom). For each of the no friction and with friction sets, progressive run numbers have increasing shake magnitudes.}
\label{fig:tableShells}
\end{figure*}

Figure \ref{fig:tableShells} lists percentages change, from initial and final stages of the simulations, of the number of smaller (yellow) particles present inside each of the ten shells. The depletion of smaller particles from the outer part of the aggregate is again clear from the negative percentages listed. It is also apparent that for Shells 1 and 2 (the innermost shells) there is little change in the number of smaller particles (and thus also larger particles) present for any of the simulation runs. We note that preliminary work by \citet{Sanchez_2010} also found a well mixed central region with a rather different simulation setup.

\subsection{The effect of friction}
\label{sec:Results:Friction}

Friction first hinders size sorting due to particle interlocking. Row 7 of Figure \ref{fig:tableShells} shows the case with friction and with a maximum shake speed of 0.92\% of the aggregate's escape speed (0.692 cm/s). Unlike its no-friction counter part (Row 1), when friction is present there are no changes in the number of smaller (yellow) particles for the smallest shake magnitude in any of the shells. When friction is present, there is a seismic activation threshold that needs to be exceeded before particles can move past each other. In the 0.92\% with friction case (Run 7), shake velocities are not large enough to overcome the frictional threshold. This suggests that there is likely a lower energy limit to impacts that are effective at triggering the Brazil Nut Effect on asteroids. 

Once the threshold is met however, friction aids in the sorting process. This is likely a result of particle ratcheting. When friction is present the uppermost shell (Shell 10) is fully depleted of smaller (yellow) particles when constituent particles are shaken at a maximum magnitude of 27.73\% of the aggregate's escape speed (20.9 cm/s) or higher. By comparison, in the no friction runs the uppermost shell is not fully depleted of smaller (yellow) particles even at the largest shake magnitude used for this study (46.21\% of the aggregate's escape speed). 

\subsection{Statistical analysis and time evolution}
\label{sec:Results:TimeEvolution}

\begin{figure*}
\includegraphics[width=1.00\textwidth]{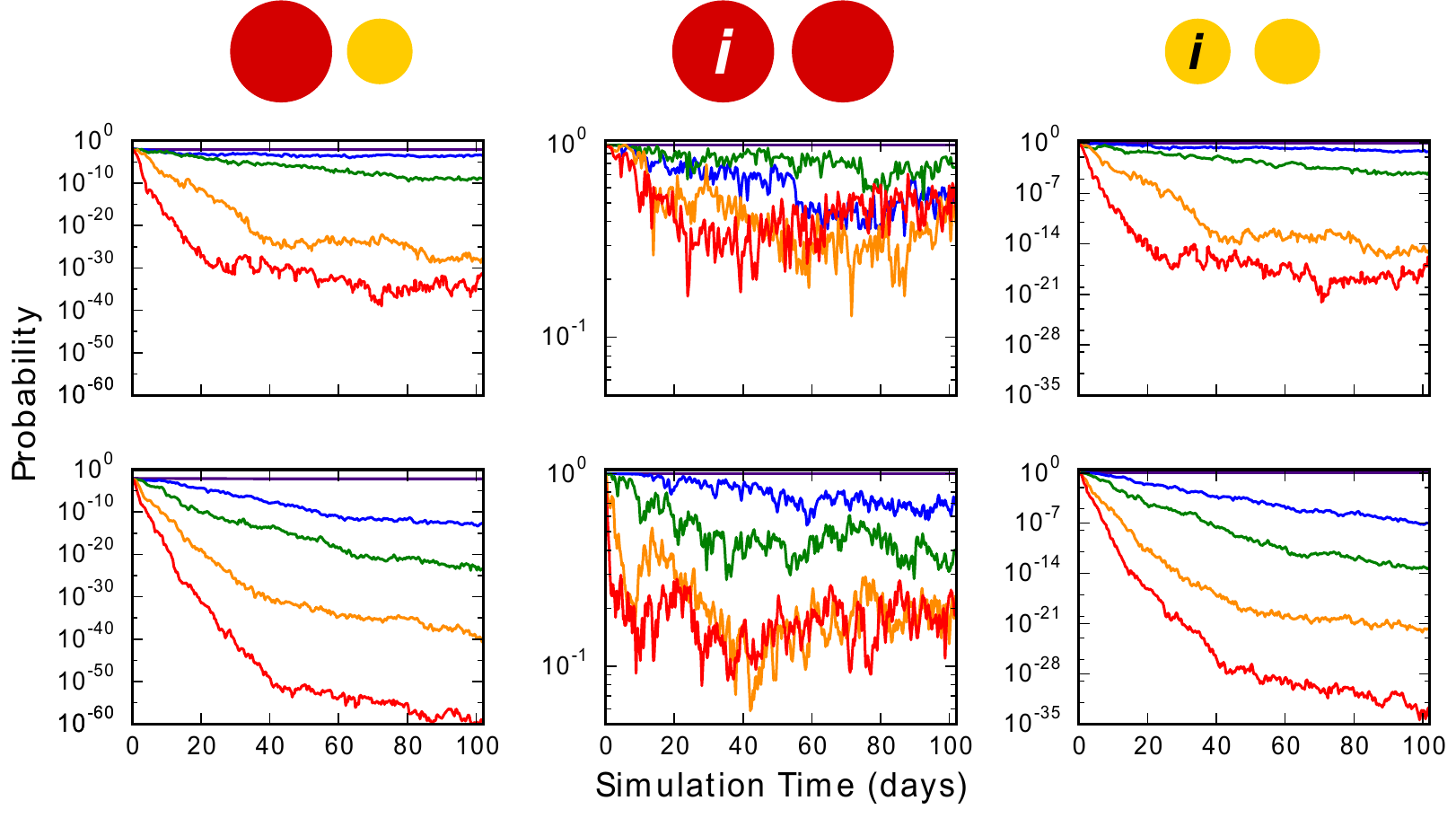}
\caption{Probabilities that particles are drawn from the same distribution as determined by the K-S test. Top row: no friction sets (Runs 1 and 3--6). Bottom row: with friction sets (Runs 7 and 9--12). Runs 2 and 8 are not shown to reduce confusion since the lines are very similar to Runs 1 and 7 respectively. Left column: probability that larger and smaller particles were drawn from the same distribution as a function of time. Middle column: comparison between the distribution of larger particles over time and the initial distribution of larger particles. Right column: comparison between the distribution of smaller particles over time and the initial distribution of smaller particles. Line colors represent maximum magnitude of shaking as a percentage of the aggregate's escape speed (indigo = 0.92\%, blue = 23.10\%, green = 27.73\%, orange = 36.97\%, and red = 46.21\%). All plots have been smoothed using a 50-point moving average.}
\label{fig:KS_simulations}
\end{figure*}

Figure \ref{fig:KS_simulations} shows K-S test results for both the no friction and friction sets. The cases with maximum speeds that are 9.24\% of the escape speed (Runs 2 and 8) are not shown as they are very similar to the 0.92\% cases. In the left-hand column we show the comparison of the large and small particle distributions, while in the center and right-hand columns we show the comparison of each of the larger and smaller distributions respectively over time with their initial distributions. Comparing the distributions over time with their initial values allows us to fully account for the minor size sorting that occurred during the formation of the aggregate ensuring we are only analyzing additional size sorting that occurred after formation. While the K-S statistic shown in the left-hand column starts from a position of significant difference between the distributions due to the size sorting during formation, these plots allow us to see whether changes in the shape of each of the individual larger and smaller particle distributions are driving the distributions to greater or lesser dissimilarity.

When considering the left column plots in Figure \ref{fig:KS_simulations}, there are several cases that show substantially increased size separation. For the no-friction set, three aggregate cases show size separation (the 27.73\% [green], 36.97\% [orange], and 46.21\% [red] cases). For the with-friction set, four aggregate cases show size separation (the 23.10\% [blue], 27.73\% [green], 36.97\% [orange], and 46.21\% [red] cases). The K-S statistic illustrates that the differences between larger particle and smaller particle distributions are highly significant in these cases. When comparing the middle column plots to the right column plots in Figure \ref{fig:KS_simulations}, they indicate that smaller particle distributions are changing significantly, while the larger particle distributions are remaining largely unchanged in shape. Even though larger particles are migrating outward, they are moving outward uniformly such that the shape of the distribution does not change dramatically while the smaller particles are filtering inwards.

\begin{figure}
\includegraphics[width=\columnwidth]{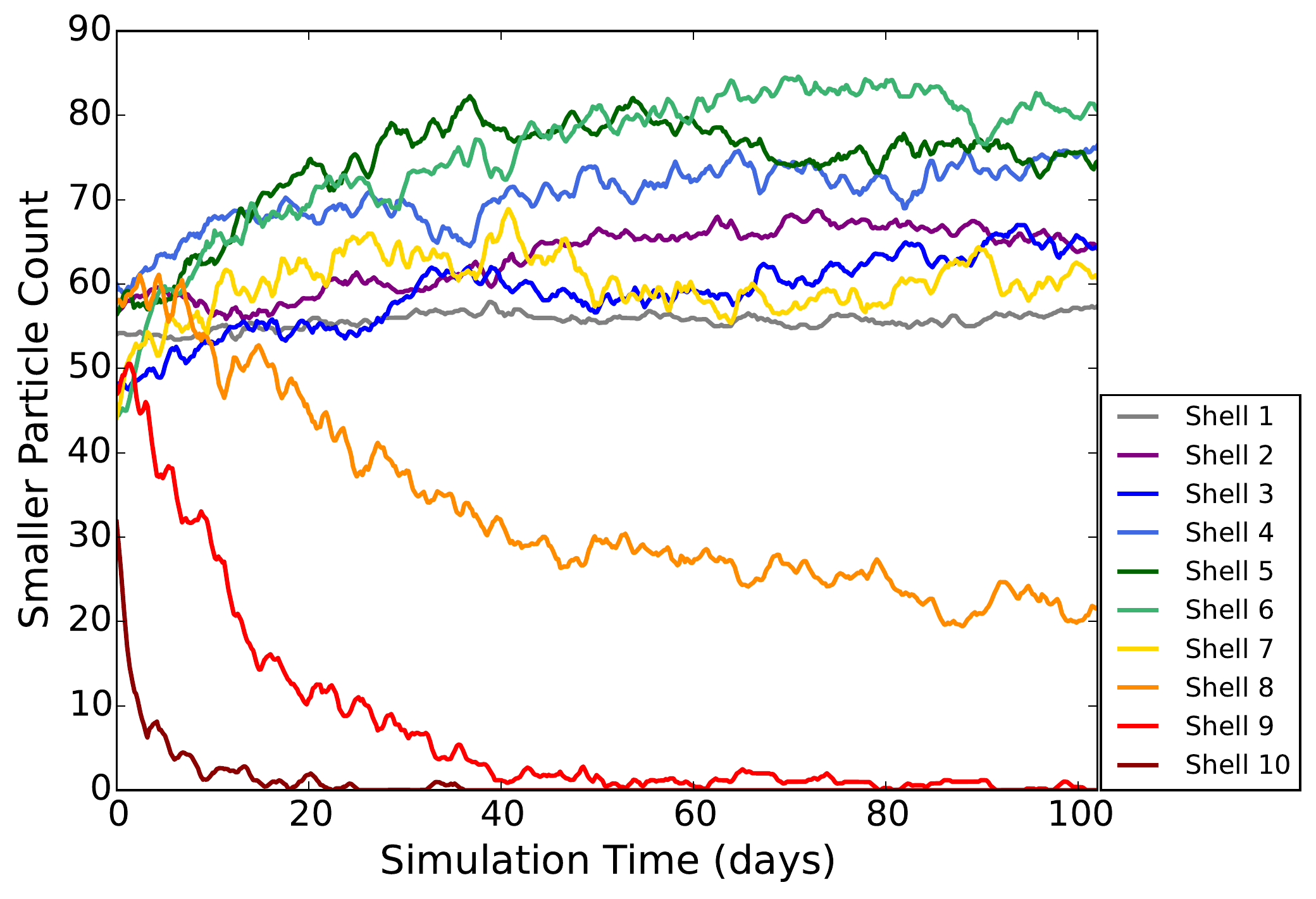}
\caption{Smaller particle count inside each of the ten shells as a function of simulation time. Shown are particle counts for the with friction case that was shaken with a maximum magnitude of 46.21\% the escape speed. All lines have been smoothed using a 100-point moving average.}
\label{fig:smalls_overtime}
\end{figure}

Previous works have shown that the timescale for the Brazil Nut Effect to take place is either proportional to $1/g$ \citep{Guttler_2013} or to $1/\sqrt{g}$ \citep{Matsumura_2014}. Though both of these works focused on the intruder model and the exact inverse factor of $g$ is uncertain, we can expect that size sorting would take longer in the interior of our aggregate (and similarly in the interiors of asteroids) due to decreasing gravity towards the center of the body. It could thus be argued that if our simulations were run for a longer period of time, even innermost regions of our aggregate would be size sorted. However, Figure \ref{fig:smalls_overtime} shows that it is not the case. In Figure \ref{fig:smalls_overtime} the number of smaller particles in each of our 10 shells over time are plotted for Run 12 (the most vigorous case with friction). Changes that occur initially can be seen to plateau off even before the simulations have reached the halfway stage. Although at lower shake speeds the simulations take longer to reach a plateau, in all cases this is still reached before the end of the simulation run. The plateauing of the number of smaller particles in each shell indicates that particles have reached an equilibrium state for the given shake speed.

\section{Discussion}
\label{sec:Discussion}

Our simulation results show that the Brazil Nut Effect occurs in our aggregates. To the extent that our simulated aggregates are representative of rubble-pile asteroids, we thus also expect the Brazil Nut Effect to occur in rubble-pile asteroids. While the effects of moving to a continuous size distribution from a binary one are not entirely clear, we would expect larger boulders to rise to the surface of an asteroid over time as the asteroid experiences impacts or other seismic shaking events. These shaking events will be subject to an activation threshold since we expect the constituents of rubble-pile asteroids to have friction, and possibly cohesion.

\subsection{Asteroid surfaces}
\label{sec:Discussion:AsteroidSurfaces}

The number of asteroids with sufficiently high resolution imaging of the surface to make any inferences about the size distribution of the surface material is rather small. One asteroid that does have such imaging is the small (535~m x 294~m x 209~m) near-Earth asteroid (25143) Itokawa. The surface of Itokawa displays regions that have substantially different ratios of larger boulders to smaller material, and \citet{Tancredi_2015} compared these with gravity maps to suggest that there is a systematic trend for lower surface gravity (`higher') regions to have greater amounts of larger boulders. They argued that this is evidence that the Brazil Nut Effect has indeed been at work on Itokawa and resulted in size sorting. In this way the irregular shape of Itokawa and other asteroids may be beneficial since the presence of regions with substantially different surface gravities can allow us to observe size sorting with only surface images, whereas for our spherical aggregates the only way to distinguish an aggregate in which the Brazil Nut Effect has brought larger material to the surface from one that is only made up of larger material is with information about the sub-surface.

While Itokawa may show evidence for size sorting it does not display complete size separation, that is regions that have larger numbers of boulders still also have finer material. There are a number of reasons why this might be the case. It is possible that the size sorting process on Itokawa has not yet had time to run to completion. While we showed that our aggregates reached a steady-state configuration in which maximal size separation had occurred, we have not attempted to match this with the expected frequency of impacts or other seismic events on rubble-pile asteroids to determine whether we would expect the Brazil Nut Effect on rubble-pile asteroids to have reached an end state. This comparison is not immediately straight forward since it is not only the number of impacts or other seismic events that matters, but how many of these exceed the activation threshold. In addition, when an asteroid undergoes a catastrophic impact and is broken up into smaller pieces, the size sorting on those fragments will likely be reset since what was formerly in the interior may now be on the surface. The time over which the Brazil Nut Effect can act is thus more likely to be the time since the last catastrophic impact rather than the age of the solar system. Detailed study of this issue is beyond the scope of this work, but we note that the Brazil Nut Effect may not have had time to reach a steady-state on all asteroids.

We must also bear in mind that an asteroid sits in the wider environment of the solar system and the seismic events that enable the Brazil Nut Effect are not occurring in isolation. Impacts that are below the seismic activation threshold, particularly micro-meteorite impacts, will gradually break up surface material into finer sizes \citep{Basilevsky_2015}, and thermal fatigue may also play a similar role \citep{Delbo_2014}. The size-distribution of material on the surface of an asteroid is thus likely to be influenced by a balance between the Brazil Nut Effect bringing larger material to the surface and other processes breaking this material down into regolith.

\subsection{The well mixed central region}
\label{sec:Discussion:CentralRegion}

A plausible explanation for why the innermost region of our aggregate is not size sorted could be that the magnitude of the shake speed was not sufficiently large. We should note, however, that our largest shake velocity, 50\% of the escape velocity, is already very large and such large shaking velocities may not be plausible in asteroids. To address this, it should be considered that asteroids are seismically shaken from their surfaces due to impacts. When impacts impart kinetic energy to their surfaces it is sufficient to influence the entire body \citep{Garcia_2015}. However, as we discussed in Section \ref{sec:Method:CompareAsteroids:ShakeModel}, due to attenuation the innermost regions will only receive some fraction of that energy, with our simulations representing the maximal case of no attenuation. Therefore, the innermost particles attaining a high velocity (larger than the 50\% of the escape speed used in our most vigorous cases) would mean the outer layers of the asteroid would have likely received velocities exceeding the escape speed. This would mean that the outer layers of an asteroid would be disrupted and removed, leaving a modified body that is smaller than the original. While some of what was previously the innermost regions will now have been size sorted in this scenario, they will now be closer to the surface of the asteroid. Therefore, we deduce that asteroids will only be size sorted in their outermost regions, retaining a well-mixed central region.  

\subsection{Examining the Driving Mechanism of the Brazil Nut Effect}
\label{sec:Discussion:DrivingMechanism}

That the distribution of larger particles remains largely unchanged while the distribution of smaller particles changes substantially is interesting and deserves closer attention. Part of this difference in the behavior of the larger and smaller particles may be a result of the greater volume occupied by the larger particles. A large particle occupies 8 times the volume of a small particle and so, clearly, when a large particle rises upwards, multiple smaller particles can move down to take its place. The precise number of small particles that can occupy the space vacated by the red particle will vary, however. The maximum packing efficiency for hexagonal close-packed spheres (of equal size) is 0.74, so within the volume of the large particle itself we could place 5--6 small particles. This neglects the voids near the original large particle however, which the smaller ones will be better able to fill, and so the removal of a large particle would generally create space for more than 5--6 small particles. The theoretical packing efficiency (of equal sized particles) is independent of particle size, and so completely replacing large particles with small particles would increase the number of particles per unit volume by a factor of 8. We can thus expect that on average each rising large particle is replaced by around 8 sinking small particles, but this will vary somewhat on a case-by-case basis.

While the difference in volume between large and small particles can account for some of the difference in the changes in the large and small particle distributions, it is unclear if it can account for all of the difference. In particular, if this was the sole reason for the difference in behavior of the evolution of the large and small particle distributions, then we would expect the large particle distribution to follow the same trend as the small particle distribution in Figure \ref{fig:KS_simulations} but with a reduced magnitude. If we consider Runs 11 and 12 (orange and red lines in the lower panels of Figure \ref{fig:KS_simulations}), we can see that although the changes in the distribution of large particles are not statistically significant, those changes that occurred do so over a much shorter time than the changes in the distribution of smaller particles. This suggests that we may require an additional factor to explain the difference in the behavior of the large and small particle distributions.

Another reason for the difference in the behavior of the large and small particle distributions may lie in the mechanism that drives the Brazil Nut Effect in these simulations. As mentioned previously, there are two mechanisms that have been postulated to mediate the Brazil Nut Effect: percolation of smaller particles through gaps created by the excitation of larger ones, and granular convection. If granular convection were the primary mechanism at work here, we would expect the distribution of large particles to undergo similarly large changes to the distribution of small particles (moderated by the greater volume of the large particles). On the other hand, if the small particles are filtering through the large ones while the large particles rise in a relatively uniform fashion, we would expect to see much smaller changes in the large particle distribution than the small particle distribution. The lack of changes in the large particle distribution could thus be an indication that percolation of the smaller particles is the primary mechanism at work in driving the appearance of the Brazil Nut Effect here.

\begin{figure*}
\includegraphics[width=1.0\textwidth]{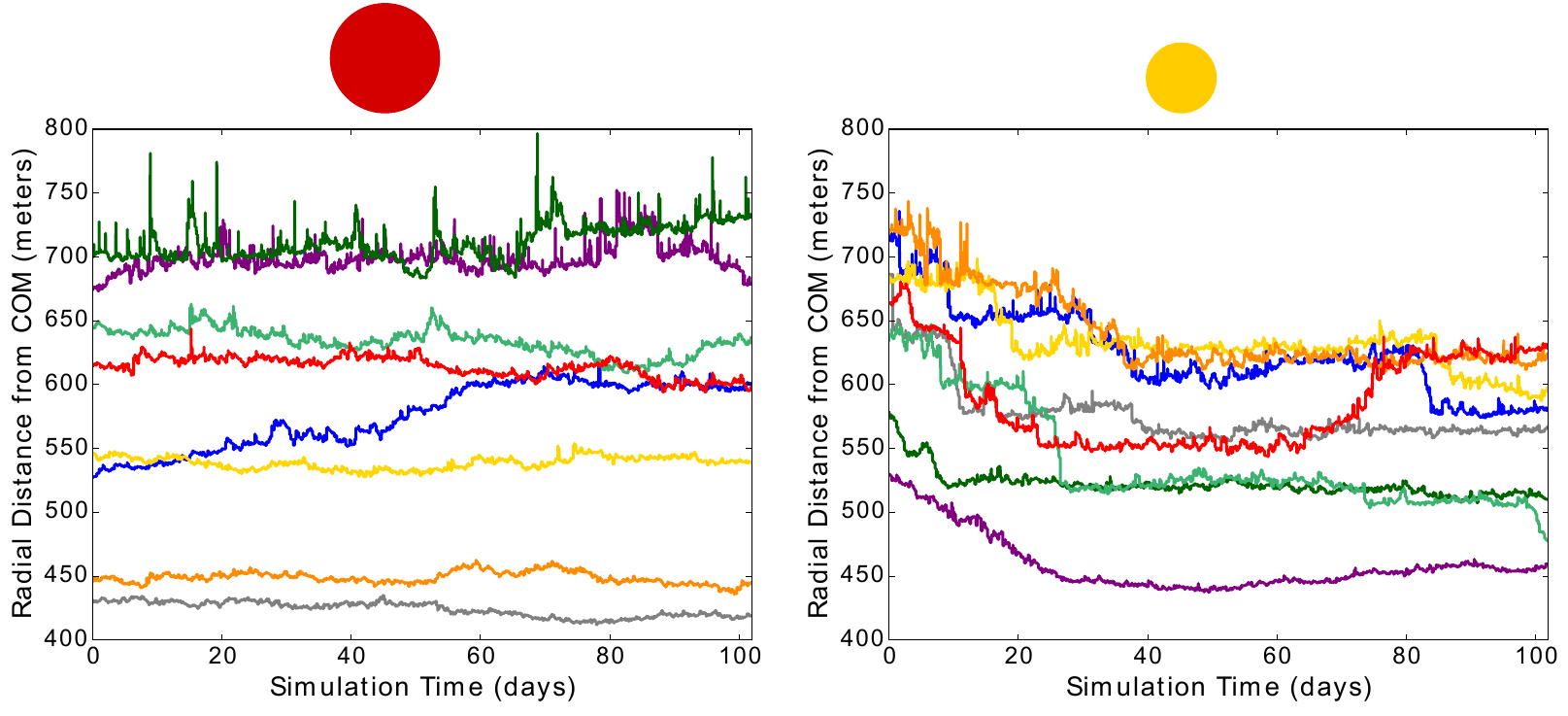}
\caption{Radial distance with respect to the center of mass (COM) as a function of time for 16 randomly selected large and small particles (8 particles each) for Run 11. Larger (red) particles are shown on the left and smaller (yellow) particles are shown on the right. Only particles that started in the outer regions of the aggregate (i.e. 400 meters from the COM or further) were selected for these plots. The colors are used to distinguish between the different randomly chosen particles.}
\label{fig:Particle_Tracker}
\end{figure*}

To examine this further we look at the motions of individual particles in detail. Figure \ref{fig:Particle_Tracker} shows the motions of 16 randomly selected particles (8 of each particle size) from the outer regions (400~m and further from the center of mass) of the aggregate in Run 11. We selected Run 11 since it has a reduced magnitude of the shakes imparted, which is shown in the plots as short, sharp upward spikes in the radial locations of the particles. Though the curves are quite noisy we can make out two behaviors in the right-hand panel of Figure \ref{fig:Particle_Tracker} (for the smaller particles): long-term oscillations in radial position and rapid drops to a new plateau level. The latter effect is the most prominent by a considerable margin, while the former is less easily discernible, but can be best seen in the red and purple curves. The left-hand panel has less evidence for any distinctive behaviors with the majority of those particles that show long-term changes in radial location showing relatively gradual rises. Long-term oscillations in location are the signature of granular convection as particles rise and fall in a convection cell. Percolation meanwhile has the signature of rapid falls inward for the smaller particles as gaps open between the larger particles allowing the smaller ones to filter down between them at stochastic intervals, while the larger particles would rise more gradually. As we believe we can see both effects, this indicates that both mechanisms are operating; however, the sudden drops account for the majority of the inward motion of the small particles (confirmed by examining many iterations of Figure \ref{fig:Particle_Tracker}). It thus appears that percolation is the dominant mechanism at work in our simulations in driving the Brazil Nut Effect.

While this analysis is suggestive that percolation is the dominant mechanism, we must note several caveats. Firstly, since as we stated in Section \ref{sec:Results:CentralRegion} the inner region of the aggregate remains well mixed, the region in which the Brazil Nut Effect occurs, and thus in which its driving mechanisms operate, is confined to the surface layers. This surface layer is relatively shallow in comparison to the size of the constituent particles of the aggregate, especially the large particles. While this is unlikely to be a hindrance to the operation of percolation, it may well inhibit the formation of convection cells and thus act to dampen granular convection. Secondly, by the same token since the surface layers in which the Brazil Nut Effect occurs are relatively shallow, we must bear in mind the problem of small number statistics.

Though there are definite indications that percolation is the primary mechanism at work in our simulations, we are cautious about applying this result to asteroids as a whole. To investigate the driving mechanisms of the Brazil Nut Effect in more detail will require a dedicated study with higher resolution simulations. For the purposes of this work however we note that this does not change our primary results; that the Brazil Nut Effect occurs in self-gravitating rubble-pile aggregates when we account for their three-dimensional shape and that the central regions remain well mixed. If granular convection is being artificially damped in our simulations due to the thinness of the surface layers, then we expect the Brazil Nut Effect should be more vigorous on asteroids. We note that if damping is due to the presence of the well mixed central region, if convection becomes more vigorous with higher resolution (smaller particle) simulations, we would not expect it to influence the well mixed central region. 

\section{Summary and Outlook}
\label{sec:Summary}

We find that in the spherical configuration the Brazil Nut Effect occurs both with and without friction. Friction hinders the sorting process at low shake velocities; however, after the frictional energy threshold is exceeded, friction works to aid the sorting process. Above a certain vibrational threshold, cases with friction require a lower shake speed to achieve the same level of size sorting as cases without friction. To the extent that our simulated aggregates are representative of rubble-pile asteroids, our results indicate that size sorting likely occurs in the outer part of rubble-pile asteroids. They also indicate however that the innermost regions should consist of a mixture of particle sizes, since even a shake magnitude of nearly 50\% of the escape speed was insufficient to sort the center. If an asteroid were to undergo an impact that resulted in the central particles acquiring a speed of 50\% of the body's escape speed, then the surface of the asteroid is likely to be disrupted.

Percolation appears to be the dominant driving mechanism behind the Brazil Nut Effect in our simulations, with granular convection playing only a minor role. We note however that the shallow depth (in terms of particle radii) of the size sorted layer due to the presence of the well mixed central region may inhibit the formation of convection cells, thus damping granular convection.

We will further explore the layers of size sorting in the future by using a larger number of particles for finer resolution. Future work will also include exploring the Brazil Nut Effect in the spherical configuration for a range of aggregate sizes, constituent particle sizes, various rotational states of aggregates, coefficients of restitution, and coefficients of friction. A spherical geometry is an idealization that we make to reduce invoking additional free parameters. Future modeling will also consider how the process would vary on bilobed asteroid shapes, and include better approximations for the input of seismic energy.


\section{Acknowledgments}

We would like to thank Derek Richardson for helping us setup \textsc{pkdgrav} and Stephen Schwartz for his advice on using the code. We would also like to thank Sanlyn Buxner for her helpful suggestions with our statistical analysis. We thank the three anonymous reviewers for their time and suggestions that improved the manuscript. The data generated by our simulations may be obtained from Viranga Perera (email: viranga@asu.edu).

\section{References}



\bibliographystyle{elsarticle-harv} 
\bibliography{PJA2015}


\end{document}